# Pressure-modulated structural and magnetic phase transitions in two-dimensional FeTe: tetragonal and hexagonal polymorphs

Wuxiao Han,[1,2] Jiajia Feng,[3] Hongliang Dong,[3] Mo Cheng,[4] Liu Yang,[5] Yunfei Yu,[1,2] Guoshuai Du,[1,2] Jiayin Li,[1,6] Yubing Du,[1,2] Tiansong Zhang,[1,2] Zhiwei Wang,[5] Bin Chen,[3] Jianping Shi,[4] and Yabin Chen[1,2,7,*]

[1]*Advanced Research Institute of Multidisciplinary Sciences, Beijing Institute of Technology (ARIMS), Beijing 100081, China.*

[2]*School of Aerospace Engineering, Beijing Institute of Technology, Beijing 100081, China.*

[3]*Center for High Pressure Science and Technology Advanced Research, Shanghai 201203, China.*

[4]*The Institute for Advanced Studies, Wuhan University, Wuhan 430072, China.*

[5]*School of Physics, Beijing Institute of Technology, Beijing 100081, China.*

[6]*School of Chemistry and Chemical Engineering, Beijing Institute of Technology, Beijing, 100081, China.*

[7]*BIT Chongqing Institute of Microelectronics and Microsystems, Chongqing, 400030, China.*

[*]Correspondence and requests for materials should be addressed to Prof. Yabin Chen: chyb0422@bit.edu.cn

# ABSTRACT

Two-dimensional (2D) Fe-chalcogenides with rich structures, magnetisms and superconductivities are highly desirable to reveal the torturous transition mechanism and explore their potential applications in spintronics and nanoelectronics. Hydrostatic pressure can effectively stimulate novel phase transitions between various ordered states and to plot the seductive phase diagram. Herein, the structural evolution and transport characteristics of 2D FeTe were systematically investigated under extreme conditions through comparing two distinct symmetries, *i.e.*, tetragonal ($t$-) and hexagonal ($h$-) FeTe. We found that 2D $t$-FeTe presented the pressure-induced transition from antiferromagnetic to ferromagnetic states at ~ 3 GPa, corresponding to the tetragonal collapse of layered structure. Contrarily, ferromagnetic order of 2D $h$-FeTe was retained up to 15 GPa, evidently confirmed by electrical transport and Raman measurements. Furthermore, the detailed *P-T* phase diagrams of both 2D t-FeTe and h-FeTe were mapped out with the delicate critical conditions. We believe our results can provide a unique platform to elaborate the extraordinary physical properties of Fe-chalcogenides and further to develop their practical applications.

# INTRODUCTION

Two-dimensional (2D) layered materials, especially transition metal dichalcogenides (TMDS), have emerged as the promising candidates in the fields of spin electronics and memory applications,[1,2] due to their exotic properties such as unconventional superconductivity,[3,4] charge density wave order,[5,6] magnetism,[7-9] and structural phase transitions.[10-12] The fascinating transformation mechanisms among the lattice structure, magnetism, and superconductivity of iron chalcogenides (*e.g.*, FeS, FeSe, and FeTe) have attracted extensive research interests nowadays.[13-18] For instance, the superconducting transition temperature of tetragonal (*t*-) FeS gradually decreases and eventually disappears with pressure increasing to ~ 4 GPa,[19] primarily originated from the pressure-induced tetragonal to hexagonal (*h*-) phase transition.[20] More interestingly, the nematic order of *t*-FeSe can change to spin density wave order at ~2 GPa, and then its magnetism is continuously suppressed till 6 GPa.[21] During compression, it is found that the continuous enhancement of superconductivity is normally accompanied with the suppressed long-rang magnetic order, indicating their complicated and ambiguous relationship.[12] When the thickness of FeSe goes to monolayer limit, the superconducting temperature has been dramatically enhanced as high as 109 K.[22] Notably, it is predicted that superconducting temperature of the intrinsic FeTe is significantly higher than that of FeSe, owing to its remarkably stronger electron-phonon coupling,[15,23-25] which needs to be proved in experiments, despite numerous studies tentatively performed on the doped FeTe.[26-30] Intriguingly, both 2D layered antiferromagnetic (AFM) *t*-FeTe and non-layered ferromagnetic (FM) *h*-FeTe nanosheets were controllably synthesized by phase-tunable chemical vapor deposition (CVD) strategy,[31-33] while the investigation on their structural and magnetic transitions is still lacking.

As the key variable in thermodynamics, hydrostatic pressure has been extensively used to nondestructively modulate the lattice structure and further tune physical properties of functional nanomaterials,[34] such as the enhanced superconductivity,[35] abnormal transport behavior,[36] and magnetic phase transitions.[37,38] It is reported that bulk *t*-FeTe exhibits rich structural transitions under pressure,[16,39-41] and various anomalies of its electrical resistance appeared with increasing pressure, partly due to the strongly entangled structural and electronic transitions.[15,42-44] Meanwhile, the magnetic studies of *t*-FeTe under hydrostatic pressure immaturely suggest the magnetic transition from low-pressure AFM to high-pressure FM order.[9,17,45] Moreover, the experimental exploration on electrical transport, structure and magnetism of *h*-FeTe under high pressure still remains blank. 2D crystalline FeTe with the tunable phases as well as magnetisms is extremely rare in TMDS family.[31,33] 2D *t*-FeTe and *h*-FeTe nanosheets with the identical chemical composition reflects the distinct lattice structures and magnetic properties, which provides us a desirable platform to explore their symmetry-magnetism correlation, and even to potentially unravel the attractive mechanism in iron-based superconductors.

In this work, we present the structural, vibrational, and electrical transport properties of 2D *t*-FeTe and *h*-FeTe nanosheets under hydrostatic compression. It is found that pressure-induced lattice collapse of 2D *t*-FeTe occurred at ~ 3 GPa, confirmed by the combined *in situ* high pressure synchrotron X-ray diffraction (XRD) and Raman spectroscopy. Meanwhile, electrical transport measurements exhibited that 2D *t*-FeTe underwent the magnetic transition from low-pressure AFM to high-pressure FM phase. The exotic anomaly in temperature-dependent resistance of 2D *h*-FeTe confirmed that FM order emerged due to electronic motion along Fe-Fe chains, well consistent with the theoretical predictions.[46] Furthermore, the detailed *P-T* phase diagrams of both 2D *h*-FeTe and *t*-FeTe were mapped out by performing numerous electrical measurements under extreme conditions. We believe these results could give assistance to understand the physical properties and transition mechanism of 2D FeTe as well as many other iron chalcogenides.

**RESULTS AND DISCUSSION**

Figure 1 shows the atomic force microscopy, high-resolution transmission electron microscopy (HRTEM), and Raman characterizations to investigate the atomic structures and lattice vibrations of 2D *t*-FeTe and *h*-FeTe crystals. The phase-tunable FeTe flakes were grown by using temperature-mediated CVD approach.[33] Atomic force microscopy results demonstrate that extracted thickness is around 5.9 nm for *t*-FeTe (Figure 1a) and 13.8 nm for *h*-FeTe (Figure 1e). The clean surface of 2D FeTe flakes is free of any cracks or contaminations. Importantly, 2D FeTe flakes obviously exhibit two distinct morphologies, 90º in *t*-FeTe and 120º (60º) in *h*-FeTe due to their own lattice symmetry, which is beneficial to intuitively distinguish phase states of any given FeTe flakes after CVD growth. To further comprehend lattice structures of 2D FeTe, the corresponding crystalline lattices are illustrated in Figure S1. Obviously, *t*-FeTe possesses a layered structure with P4/nmm space group, in which Fe atoms are distributed between the double slabs of Te atomic layers along the interlayer direction. In comparison, *h*-FeTe belongs to P6$_3$/mmc space group with a non-layered structure.

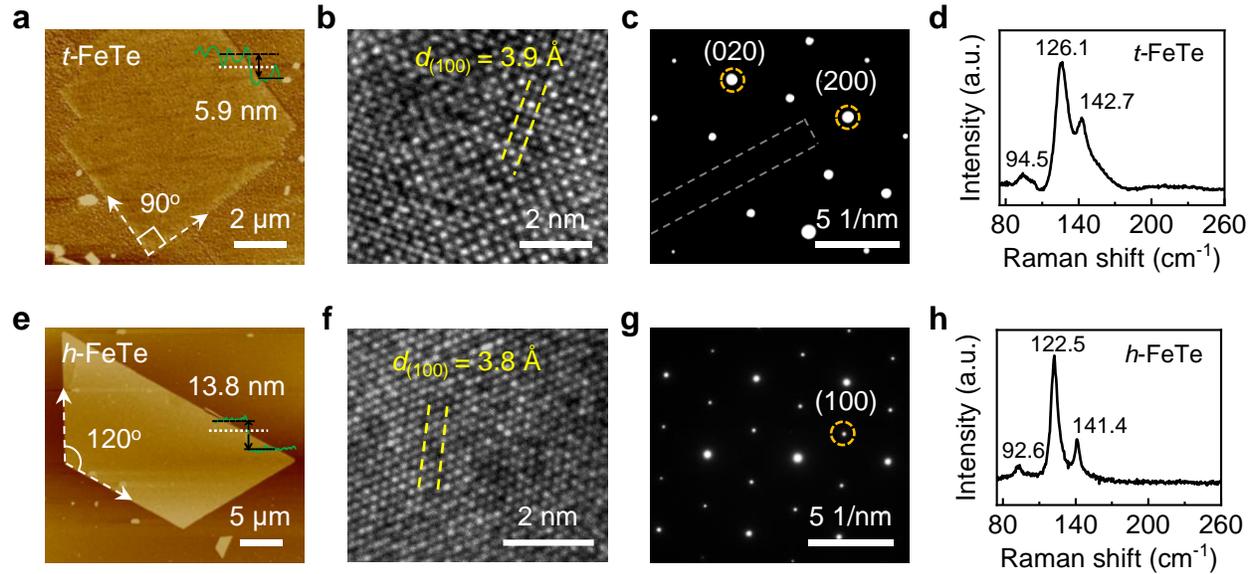

**Figure 1. Morphological and structural characterizations of 2D crystalline *t*-FeTe and *h*-FeTe nanosheets.** (a, e) Atomic force microscopy images of *t*-FeTe nanosheet (a) with a thickness of 5.9 nm and *h*-FeTe nanosheet (e) with a thickness of 13.8 nm. The dashed lines with arrows feature the characteristic lattice angle in each FeTe phases. (b, f) Representative HRTEM images of *t*-FeTe (b) and *h*-FeTe (f) nanosheets, and their distances between (100) planes are labeled. (c, g) The corresponding SAED patterns of *t*-FeTe (c) and *h*-FeTe (g) crystals. The sharp diffraction spots indicate high quality of the crystalline FeTe samples. The gray dashed box indicates the TEM baffle which potentially obscures some diffraction signals. (d, h) The typical Raman spectra of *t*-FeTe (d) and *h*-FeTe (h) nanosheets.

To investigate the atomic structures of 2D FeTe nanosheets, the detailed HRTEM measurements were carried out along [001] zone axis, as presented in Figure 1b and 1f. The obtained lattice constant 3.9 Å of *t*-FeTe is relatively larger than that (3.8 Å) of *h*-FeTe, as indicated by the yellow dashed lines. It is clear that atom arrangements of 2D FeTe perfectly comply with their specific lattice symmetries, which is further confirmed by selected area electron diffraction (SAED) measurements. The SAED patterns display the four- and six-folded symmetry in Figure 1c and 1g, respectively, and the calculated lattice constants are quantitatively consistent with HRTEM results. Meanwhile, the bright and sharp diffraction spots suggest the high quality of our 2D crystalline FeTe. We further performed Raman measurements to investigate lattice vibrations of 2D FeTe flakes. The Raman shifts at 126.1 and 142.7 cm$^{-1}$ of $t$-FeTe in Figure 1d are slightly larger than those (122.5 and 141.4 cm$^{-1}$) of $h$-FeTe flake in Figure 1h, which can be assigned to $E_g$ and $A_{1g}$ modes, respectively.[31,33] According to group theory, $E_g$ mode vibrates along in-plane direction, while $A_{1g}$ mode along out-of-plane direction.

To explore the physical insights into pressure-effect on structural transition of 2D FeTe, we performed *in situ* Raman measurements on both $t$-FeTe and $h$-FeTe up to ~ 8.5 GPa. 2D FeTe nanosheets were transferred onto diamond surface by polystyrene-assisted method (Figure S1).[33] Figure 2a presents the representative Raman spectrum of a $t$-FeTe nanosheet (25.8 nm thick). As pressure increases, Raman shift of $E_g$ mode displays a significant redshift, owing to the softened phonon vibration. As shown in Figure 2c, the extracted wavenumber by Lorentz fitting decreases monotonically (linear slop ~ -2.2 cm$^{-1}$/GPa) during compression, while it exhibits a remarkable hysteresis at ~ 3 GPa during decompression. Contrarily, it is found that $A_{1g}$ mode is highly sensitive to pressure, and its frequency blueshifts firstly and then slightly redshifts before ~ 3 GPa, followed by a hardening behavior once again. Raman intensity of $E_g$ and $A_{1g}$ modes become undetectable up to ~ 8 GPa and come back upon decompression, suggesting the reversible transition of $t$-FeTe. As displayed in Figure 2b and 2d, pressure-effect on Raman spectrum of 2D $h$-FeTe (thickness 25.4 nm) show the comparable phenomena to $t$-FeTe, indicating that these phonon modes are not evidently affected by lattice symmetry. Notably, both $E_g$ and $A_{1g}$ modes in FeTe originate from the vibration of Te atoms. The weakened $E_g$ modes of $t$-FeTe and $h$-FeTe under pressure are primarily attributed to the enhanced electron-phonon coupling, which is confirmed in other typical telluride materials.[11,47,48]

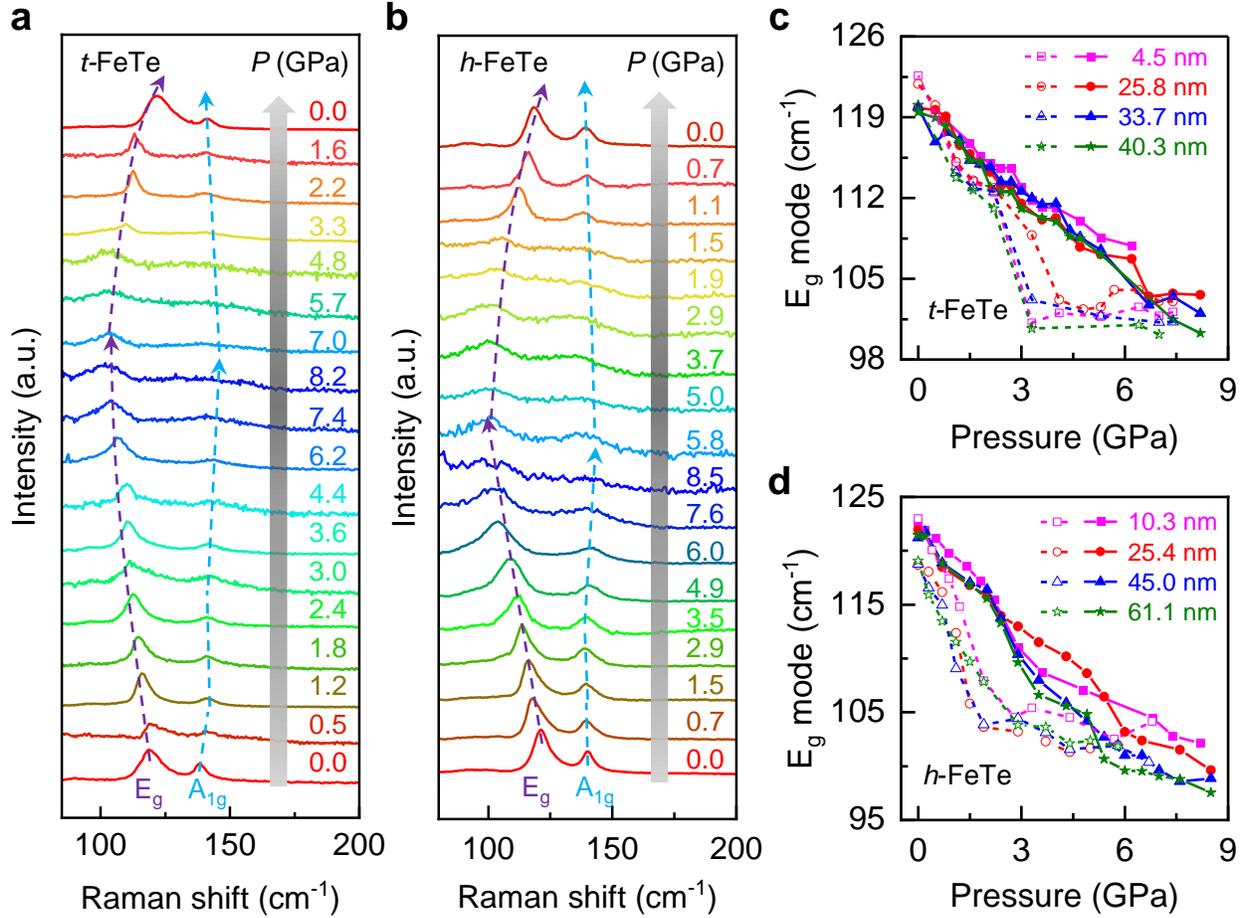

**Figure 2. Pressure-dependent Raman spectrum of 2D *t*-FeTe and *h*-FeTe with variable thickness.** (a, b) Raman spectrum of 25.8 nm-thick *t*-FeTe (a) and 25.4 nm-thick *h*-FeTe (b) under various pressures up to ~ 8 GPa. The dash lines with arrows are drawn as guide for the eyes. (c, d) Pressure-dependent Raman shift of $E_g$ mode of *t*-FeTe (c) and *h*-FeTe (d) with different thicknesses. The solid and hollow symbols denote Raman shifts of $E_g$ mode under compression and decompression process, respectively.

Furthermore, we acquired Raman spectrum based on many *t*-FeTe and *h*-FeTe nanoflakes with the variable thickness from ~4.5 to 120 nm, and their pressure response exhibits the profound thickness-dependence (more details in Supplementary Figures S2-S4). As shown in Figure 3, the compressibility $K$ of in-plane $E_g$ mode of *t*-FeTe, defined as $\partial\omega/\partial P$, slightly increases from -2.0 to -2.3 cm$^{-1}$/GPa, corresponding to the varied thickness from 4.5 to 69.6 nm. Moreover, *h*-FeTe presents the stronger compressive behavior, and the representative compressibility is -2.6 and -3.3 cm$^{-1}$/GPa for 10.3 and 119.4 nm-thick nanoflakes. Importantly, for thicker FeTe nanoflakes, their slope $\partial\omega/\partial P$ almost remains constant and approaches to the bulk limit when the thickness exceeds ~ 45 nm. In principle, the Raman shift rate with pressure can be related to the force constant $k$ of this phonon vibration, and partly reflects the elastic modulus of FeTe nanoflakes, according to dynamic theory of crystal lattices.[49] The force constant $k$ principally decreases as the thickness of FeTe nanosheets before the emergent plateau, indicating the gradually reduced Young's modulus of FeTe along in-plane direction. More interestingly, the slope $\partial\omega/\partial P$ of $t$-FeTe nanosheets is dramatically greater than $h$-FeTe nanosheets, potentially due to the larger elastic modulus and layered structure of $t$-FeTe.

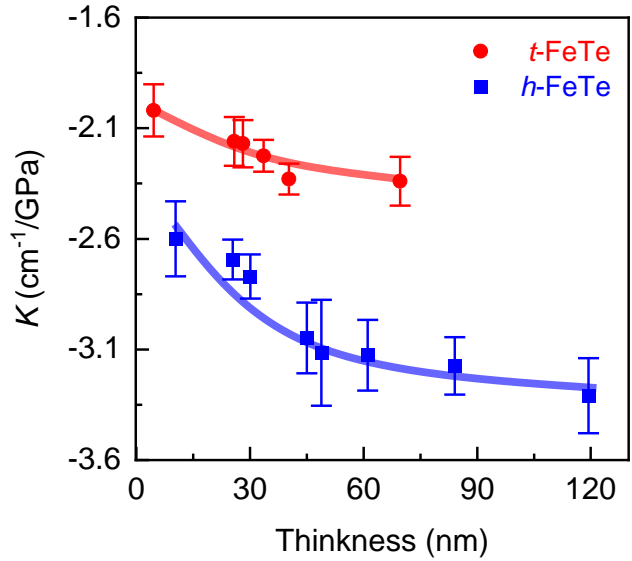

**Figure 3. The slope $K\sim\partial\omega/\partial P$ of $E_g$ mode varies with thicknesses of 2D FeTe.** The compressibility of $t$-FeTe (red) is apparently smaller than that of $h$-FeTe (blue) at a given thickness. The uncertainty of $K$ from linear fitting is considered as well. The red and blue lines are guide for the eyes.

To prove the above hypothesis, nanoindentation experiments based on atomic force microscopy were performed to measure the elastic module of 2D FeTe nanosheets, as shown in Figure S5. This advanced technique has been maturely established to study mechanical properties of numerous 2D layered nanomaterials, such as graphene and black phosphorus.[50-52]

2D *t*-FeTe and *h*-FeTe nanosheets were transferred onto a silicon substrate with pre-fabricated ~ 1.5 μm circular grooves. The morphological characterizations of *t*-FeTe and *h*-FeTe nanosheets revealed their clean and flat surface without any bubble or wrinkle. The load-indentation depth curve can be reasonably fitted with the following equation: $F = (\sigma_0^{2D}\pi)\delta + (E^{2D}\frac{q^3}{r^2})\delta^3$, where $F$ is the applied load, $\sigma_0^{2D}$ is the pretension force, $\delta$ is the indentation depth, $r$ is the radius of the hole, $E$ is the elastic modulus, and $q$ is a dimensionless constant calculated by the equation $q = 1/(1.05 - 0.15\upsilon - 0.16\upsilon^2)$. Poisson's ratio $v$ of FeTe is taken as ~ 0.21 from literature.[53] Finally, the derived Young's modulus of 2D *t*-FeTe approximates to 117.0 ± 7.8 GPa, which is obviously larger than 96.4 ± 5.4 GPa of *h*-FeTe. This result totally agrees with the Raman data as discussed previously.

In order to probe its structural stability and phase transitions, *in situ* synchrotron XRD measurements of 2D *t*-FeTe were explored under high pressure up to 37.3 GPa. Figure 4a shows the representative XRD results, of which the diffraction peaks can be well indexed with the tetragonal (space group P4/nmm) structure. As pressure increases, all diffraction angles became larger and no new peak appeared. Figure 4b exhibits the detailed Rietveld refinements at 2.9 GPa with the accepted R-weighted pattern $R_{wp}$ as low as 5.7%. Furthermore, the lattice constant *c* shrinks by ~ 10% under 37.3 GPa, evidently larger than 8% of *a* or *b*, owing to the layered characteristic of *t*-FeTe. The van der Waals interaction along *c* direction is significantly weaker than the intralayer covalent bonds, which occurred in FeSe and FeS as well.[20,54] The continuous *P-V* curve without any kink suggests no phase transition happened. Notably, *c* is rapidly reduced by 0.2 Å within 3.9 GPa, followed by the gradual decrease. This unexpected phenomenon can be interpreted as a lattice collapse in *t*-FeTe, consistent with collapsed tetragonal phase of iron telluride, where spin state of iron plays a key role to induce its lattice distortion under pressure.[55]

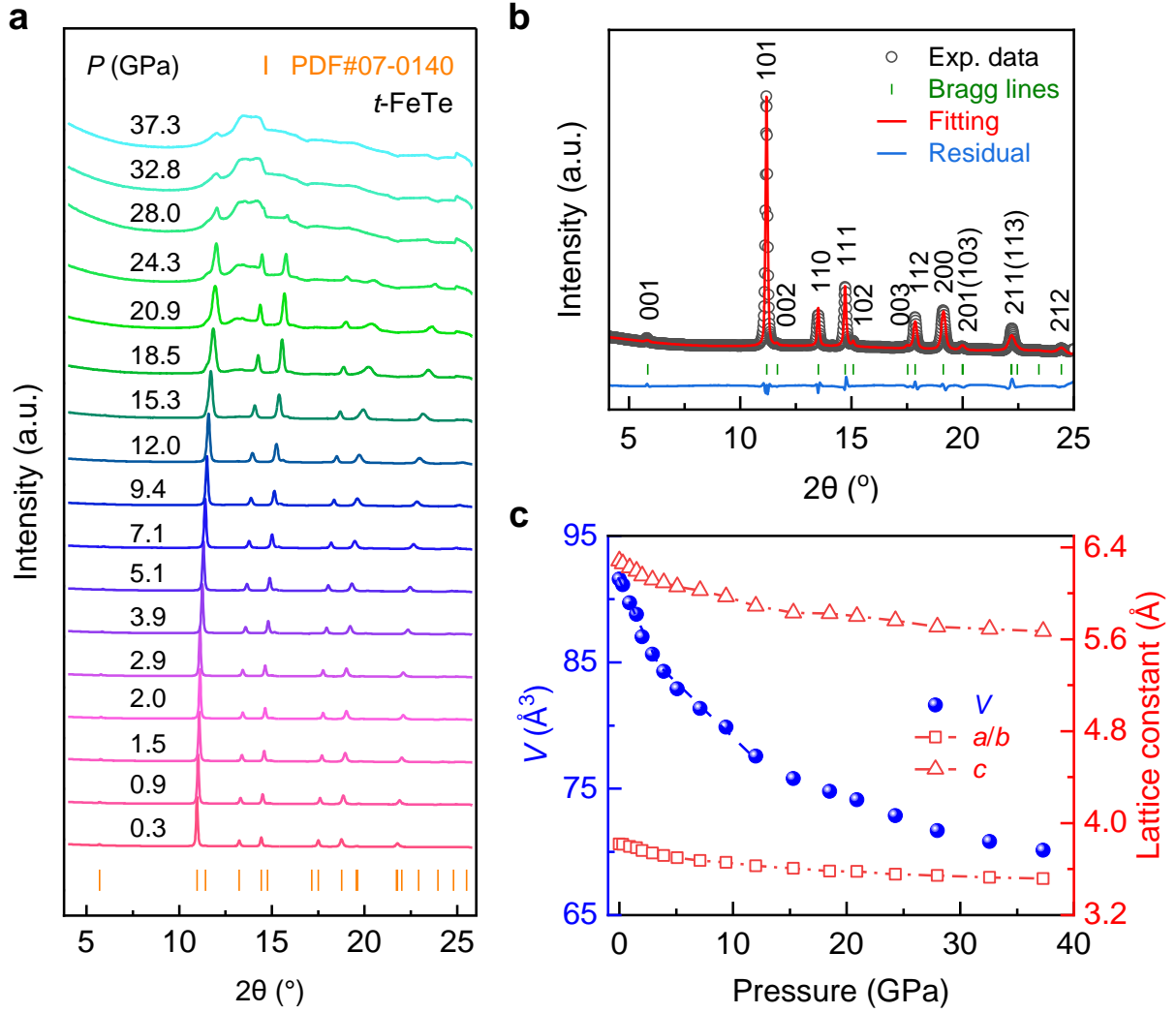

**Figure 4. *In situ* synchrotron XRD measurements of *t*-FeTe under various pressures.** (a) the acquired synchrotron XRD curves of *t*-FeTe as pressure increases up to 37.3 GPa. The diffraction peaks are in good agreement with the standard (PDF#07-0140) (b) Rietveld refinements of the representative XRD curve of *t*-FeTe at 2.9 GPa. Black cycles represent the experimental data, red line means the fitting result after refinement, green vertical bars represent the standard Bragg peaks, and blue curve denotes the difference between experimental and calculated results. (c) Pressure-dependence of the volume *V* (blue) and lattice constants *a*, *b* and *c* (red) of *t*-FeTe.

The *P-V* data in Figure 4c can be well fitted using third-order Birch-Murnaghan equation of state: $P(V) = \frac{3}{2}B_0 \left[\left(\frac{V_0}{V}\right)^{\frac{7}{3}} - \left(\frac{V_0}{V}\right)^{\frac{5}{3}}\right] \left\{1 + \frac{3}{4}(B_0' - 4)\left[\left(\frac{V_0}{V}\right)^{\frac{2}{3}} - 1\right]\right\}$, where $V_0$ and $V$ denote the unit cell volume under ambient pressure and high pressure, respectively. $B_0$ is bulk modulus and $B_0'$ is the pressure derivative of $B_0$ evaluated at ambient pressure. When $B_0'$ was fixed as 4, $B_0 = 35.5 \pm 3.8$ GPa and $V_0 = 91.8 \pm 0.3$ Å$^3$ were obtained for $t$-FeTe under the pressure up to 3.9 GPa, while $B_0 = 70.4 \pm 5.0$ GPa and $V_0 = 88.6 \pm 0.4$ Å$^3$ were extracted for collapsed $t$-FeTe within 4 – 10 GPa. We did not find the tetragonal to hexagonal transition in 2D FeTe, which normally occurred in FeS and FeSe.[19,21] This unexpected result of 2D FeTe can be mostly ascribed to the larger radius of Te anion than S and Se, implying its stronger interlayer interaction and thus higher transition pressure.

To explore the transport characteristics of 2D FeTe, four-terminal nanodevices were demonstrated to probe its magnetic and electrical properties. Figures 5a and 5c shows the temperature-dependent resistance $R(T)$ results of $t$-FeTe (30.1 nm thick) and $h$-FeTe (50.6 nm thick) under one specific pressure, and the maximum pressure reached ~ 9.8 and 13.6 GPa, respectively (more details and analysis shown in Figures S6 and S7). In Figure 5a, it is obvious that the $R$-$T$ slope at a given pressure switches from positive to negative value, corresponding to the magnetic ordering transition of $t$-FeTe. At low pressure regime ($P < \sim 2.7$ GPa), the slope change is very apparent, resulted from paramagnetic (PM) to AFM transition of $t$-FeTe at its Néel temperature ($T_N$).[45] Under ambient pressure, $R$-$T$ data displays anomalous around $T_N \sim 65.7$ K in Figure S8, well consistent with literature.[15,31] As exhibited in the mapped $P$-$T$ diagram (Figure 5b), the extracted $T_N$ significantly decreases as pressure, and approaches its minimum of ~ 44 K at ~ 2.7 GPa, suggesting the gradually weakened AFM state. At the mid-pressure range, the manifest change in $R$ vs $T$ data originated from the commensurate AFM to FM order of $t$-FeTe, in accordance with both the muon and neutron results.[56] In addition, when pressure exceeded ~ 3.4 GPa, the FM order of $t$-FeTe transformed to PM state as temperature rose up, and the corresponding Curie temperature ($T_C$) keeps almost unchanged with pressure. Notably, the pressure-dependent resistance of $t$-FeTe at 2 K subtly decreased with pressure in Figure S8, implying a complete AFM to FM phase transition at ~ 4.8 GPa.

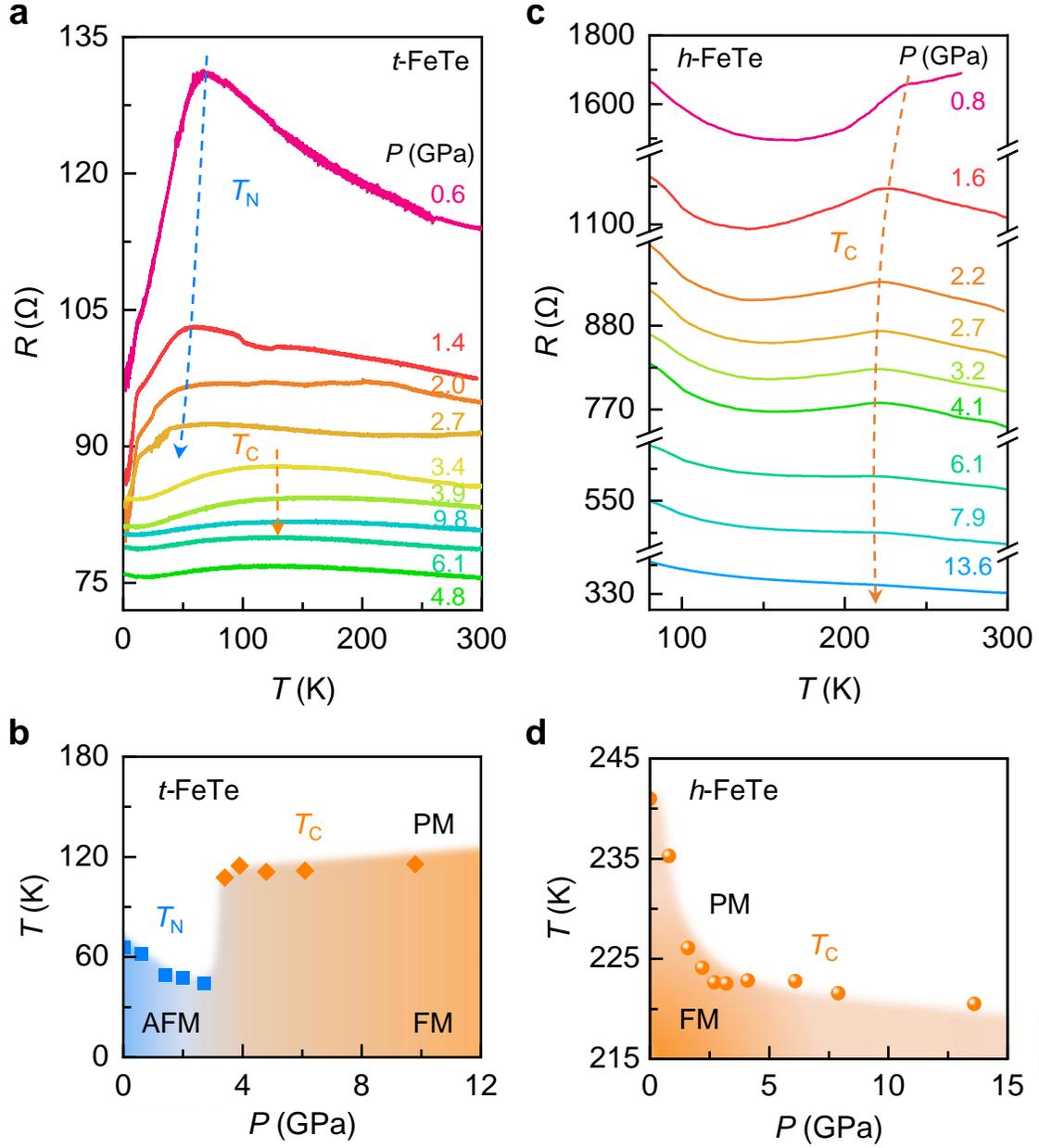

**Figure 5. Evolution of temperature-dependent resistance with pressure and the plotted temperature-pressure magnetic phase diagram of 2D FeTe.** (a, c) Temperature-dependent resistance of 2D *t*-FeTe (a) and *h*-FeTe (c) nanodevices under different pressures, and their thickness is 30.1 and 50.6 nm, respectively. The dashed lines indicate the trend of transition temperature. (b) *P-T* phase diagram of 2D *t*-FeTe with different magnetic states. The AFM *t*-FeTe transforms to FM phase as pressure increases at low temperature. (d) *P-T* phase diagram of 2D *h*-FeTe. Notably, $T_C$ decreases dramatically at low pressure range.

In contrast, 2D *h*-FeTe nanoflakes present the intrinsic FM order at low temperature and ambient pressure, different from AFM *t*-FeTe. The temperature-dependent *R*(*T*) of *h*-FeTe (thickness ~ 38.2 nm) displayed a remarkable slope change at ambient pressure as shown in Figure S8, corresponding to its PM to FM ordering transition.[31] The extracted $T_C$ data from *R*-*T* curves under variable pressures (Figure 5c) allows us to plot the fine phase diagram covering FM and PM regions, as presented in Figure 5d. FM state of *h*-FeTe is evidently weakened as pressure increasing to ~3 GPa, accompanied with the significantly reduced $T_C$ from 241 (0 GPa) to 222 K. When *P* goes up further, the transition temperature remains the similar level till the maximum pressure of ~ 13.6 GPa. Furthermore, the overall resistance of *h*-FeTe always decreases gradually with pressure as shown in Figure S9, unlike to the complex behavior in *t*-FeTe.

**CONCLUSION**

In conclusion, we used 2D FeTe as the representative Fe-chalcogenide to explore its lattice symmetry-dependent phonon vibrational, mechanical, and electrical properties under extreme conditions, by rationally comparing the *t*-FeTe and *h*-FeTe nanoflakes. Intriguing, We found that 2D *t*-FeTe exhibited pressure-induced structural collapse of tetragonal lattice at ~ 3GPa, accompanied with the significant AFM to FM transition proved by Raman spectroscopy, synchrotron XRD, and electrical transport measurements. The *R*(*T*) results confirmed that 2D *h*-FeTe with FM order was reserved without any apparent transition. The plotted full *P*-T phase diagrams for 2D *t*-FeTe and *h*-FeTe shows the remarkable divergence due to their distinct lattice symmetries. The results in this study can shed light to discover more exotic phases in Fe-chalcogenides and hence to investigate their fantastic physical properties.

**EXPERIMENTAL METHODS**

**Preparation and Characterization of 2D FeTe Samples**

High-quality FeTe single crystals with two different phases were synthesized on mica substrate by an atmospheric pressure CVD approach. Details of the sample preparation can be found in the previous work.[33] The optical images of FeTe nanosheets were obtained by optical microscopy (Olympus BX53M). Atomic force microscopy (Dimension FastScan, Bruker) characterizations were performed with tapping mode. For HRTEM measurements (Tecnai TF-20), FeTe nanosheets were transferred from mica substrate onto TEM grids by using the non-polar polystyrene-assistant method, and the detailed procedure can be found elsewhere.[33] The accelerating voltage was as low as 80 kV.

***In Situ* Raman Measurements**

*In situ* Raman measurements under high pressure was carried out by using our home-made Raman system, equipped with iHR550 spectrometer and 633 nm laser as the excitation source. The laser power was as low as tens of μW to exclude overheating effect. A symmetric diamond anvil cell (DAC) was used to generate hydrostatic pressure. T301 gasket was pre-indented to ~ 50 μm in thickness, and further drilled to give a center hole of ~ 120 μm as sample chamber. The 2D FeTe nanosheets with different thickness were transferred from mica surface onto diamond culet. Silicon oil was used as pressure-transmitting medium (PTM) to offer the hydrostatic pressure condition. The pressure was calibrated by the ruby fluorescence method.

**High-Pressure XRD Characterizations**

For *in situ* XRD measurement, rhenium gasket was utilized, owing to its superior hardness. The well-grinded and pre-compressed *t*-FeTe samples were loaded into the DAC chamber, and silicone oil was used as PTM. The XRD patterns were collected at the 4W2 High Pressure Station with a Pilatus detector in the Beijing Synchrotron Radiation Facility (BSRF). The wavelength of focused monochromatic X-ray beam was 0.6199 Å. High-purity $CeO_2$ sample was used as the internal standard for detector calibration. The collected 2D diffraction patterns were converted to the intensity versus $2\theta$ curves by Dioptas software, and then analyzed by Rietveld method using GSAS package.[57]

**Electrical Measurements under High Pressure**

For *in situ* electrical measurements, a non-magnetic Be-Cu DAC was selected to apply for high pressure. The device fabrication process of FeTe nanosheets is summarized in Figure S6. The 2D FeTe nanosheets were conveniently transferred to diamond culet by the dry-transfer method as described above. Then, one inorganic nanowire with diameter of ~ 5 μm was transferred on FeTe nanosheet as hard mask for e-beam evaporation. 10 nm Ti/100 nm Au films were deposited as metallic electrodes. A thin *h*-BN flake as protective layer was transferred to cover the entire nanodevice, as shown in Figure S6. The powder of *c*-BN/epoxy mixture was used to electrically insulate FeTe from metallic gasket. Platinum foils as the outer electrical lead contacted with copper wires by silver glue. Daphene 7373 was used as the PTM, because of the produced quasi-hydrostatic condition even under low temperature. Pressure was measured via the ruby fluorescence method before and after each cooling process. DAC was placed inside the OptiCool system (Quantum Design) to perform the temperature-dependent experiments from 300 down to 2 K. The electrical resistance was acquired through the 2182A nanovoltmeter and 6221 current source (Keithley).

## ASSOCIATED CONTENT

**Supporting Information**

The Supporting Information is available free of charge and can be found online.

Schematics for the high-pressure DAC setup and fabrication method of four-terminal nanodevice; Pressure-dependent Raman spectrum of *t*-FeTe and *h*-FeTe with various thickness; $A_{1g}$ mode variation of 2D FeTe under different pressure; Young's modulus measurements of 2D FeTe by nanoindentation approach based on atomic force microscopy; Electrical transport of 2D *t*-FeTe and *h*-FeTe under variable temperature and pressure.

**Notes**

The authors declare no competing financial interests.

**Data Availability**

All data related to this study are available from the corresponding author on reasonable request.

**Author Contributions**

Y.C. and W.H. conceived this research project and designed the experiment. J.F., H.D., and B.C. provided the technical support on electrical measurements and XRD characterizations. M.C., Z.W., and J.S. synthesized the FeTe samples. W.H., Y.Y., G.D., J.L., T.Z., and Y.D. performed the structural characterizations of FeTe by Raman spectroscopy and HRTEM. W.H. and Y.C. wrote the manuscript with the necessary input of all authors. All authors have given approval of the final manuscript.

**Acknowledgements**

This work was financially supported by the National Natural Science Foundation of China (grant numbers 52072032, 92164103, and 12090031) and the 173-JCJQ program (grant No. 2021-JCJQ-JJ-0159).